\magnification=1200 \baselineskip=13pt \hsize=16.5 true cm \vsize=20 true cm
\def\parG{\vskip 10pt} \font\bbold=cmbx10 scaled\magstep2

\centerline{short version in {\it Theory in Biosciences}
{\bf 120}(1) (2001)}\parG\parG

\centerline{\bbold Why do Evolutionary Systems}\parG
\centerline{\bbold Stick to the Edge of Chaos}\parG
\centerline{Paulo Murilo Castro de Oliveira$^{\rm \dag}$}\parG
Laboratoire de Physique et M\'ecanique des Milieux H\'et\'erog\`enes\par
\'Ecole Sup\'erieure de Physique et de Chimie Industrielles de la Ville
de Paris\par
10, rue Vauquelin, 75231 Paris Cedex 05, France\par

\vskip 0.4cm\leftskip=1cm\rightskip=1cm 

{\bf Abstract}\parG

        The long-term behaviour of dynamic systems can be classified in
two different regimes, regular or chaotic, depending on the values of the
control parameters, which are kept constant during the time evolution.
Starting from slightly different initial conditions, a regular system
converges to the same final trajectory, whereas a chaotic system follows
two distinct trajectories exponentially diverging from each other.\par

        In spite of these differences, regular and chaotic systems share a
common property: both arrive exponentially fast to their final destiny,
becoming trapped there. In both cases one has finite transient times. This
is not a profitable property in what concerns evolutionary strategies,
where the eternal search for new forms, better than the current one, is
imperative. That is why evolutionary dynamic systems tend to tune
themselves in very particular situations in between regular and chaotic
regimes. These particular situations present eternal transients, and the
system actually never reaches its final destiny, preserving diversity.
This feature allows the system to visit other regions of the space of
possibilities, not only the tiny region covered by its final
attractor.\par

\leftskip=0pt\rightskip=0pt\vskip 0.4cm

\par

\vfill
$^{\rm \dag}$ permanent address:\par
Instituto de F\'\i sica, Universidade Federal Fluminense\par
av. Litor\^anea s/n, Boa Viagem, Niter\'oi, Rio de Janeiro, Brasil
24210-340\par
PMCO @ IF.UFF.BR

\eject

{\bf \item{I)} Introduction}\parG

        Any dynamic system has its dynamic variables which evolve as time
goes by, and also its constant parameters. An example is the so-called
logistic map

$$x_{t+1}\, =\, a\, x_t (1 - x_t)\,\,\,\, ,\eqno(1)$$

\noindent where $x_t$ is the dynamic variable which follows the trajectory
$x_0, x_1, x_2 \dots$ after starting from the initial seed $x_0$. The time
$t = 0, 1, 2 \dots$ is discrete. The control parameter $a$ is kept
constant during the dynamic evolution. This map was introduced a century
and a half ago in order to model population growth dynamics [1], by
considering $x_t = P_t/P_{\rm max}$, where $P_t$ is the current population
at time (or generation) $t$, and $P_{\rm max}$ is the maximum population
the environment can support. Within this model, $a$ represents the birth
rate per generation, allowing population growth if $a > 1$ for small
values of $x_t$ (i.e. $P_t <<< P_{\rm max}$). The Verhulst death factor
$(1 - x_t)$ avoids population explosion whenever $P_t$ approaches the
maximum environment capacity $P_{\rm max}$, provided $a \le 4$.

        Other dynamic systems cannot be put in a simple, compact analytic
language as in equation (1). Nevertheless, the dynamic variable as well as
the rules governing its time evolution exist, and must be described
according to some other language. For instance, the genetic pool of a real
population evolves in time, but cannot be expressed by a single number
$x_t$. At least one needs to count the frequency of each possible allele
for all genes, among the current population at time (or generation) $t$,
storing the results on a one-dimensional array of frequencies. In this
case, instead of a number, the dynamic variable is this array. In order to
consider possible correlations between these genes, for instance how many
individuals share the same set of $D$ alleles, the frequencies must be
stored on a $D$-dimensional array which becomes the dynamic variable.
Instead of frequencies, the most complete representation is to store, for
each individual, a bit 1 or 0 for each possible present or absent allele:
in this case, the current populations is represented by a variable number
of bit-arrays, or bit-strings, one for each alive individual, i.e. the
dynamic variable becomes a $P$-dimensional array of bit-strings, where the
population size $P$ varies. Also the rules allowing to determine the
current genetic status of the population from the knowledge of its past
history depends on its reproduction and death behaviour, environment
changes, and so on. Hardly these rules can be translated into equations
relating the current and previous arrays. Perhaps other more detailed
genetic information cannot even be stored in arrays of numbers, and the
mathematical description of the system becomes very difficult. Many other
non-biological evolutionary systems, as the behaviour of an economy with
its various individual agents, or the development of a human language,
fall into the same class of problems, with more or less the same
difficulties.

        A very powerful language to express the rules governing the
evolution of such a system, is computer modeling. Let's consider a diploid
population, as a simple example. Each individual is represented by two
$B$-bit-strings storing, say, $B = 1024$ bits each. These two ordered
$B$-strings will be called the individual's genome, kept fixed during its
whole life. It inherits one $B$-string from its mother, and the other from
its father. First, before its birth, the mother's genome is copied, and a
random crossover is performed on the copy: both $B$-strings are cut in the
same random position, and a new $B$-string is constructed by joining one
random piece with the complementary piece taken from the other string.
This is the $B$-string inherited from the mother, and the other is
constructed by performing the same process on the father's genome. Then,
$M$ random mutations are introduced, say $M = 1$, by flipping $M$ randomly
chosen bits (from 0 to 1, or vice-versa) of the newborn genome. Besides
its genome, each newborn also inherits the mother's family name, without
mutations. During each time step, each individual reproduces $F$ times as
mother, say $F = 1$. This breeding process is sequentially performed for
each alive individual, with the father being another one, randomly chosen.

        The population size $P$ (the number of alive individuals) is kept
nearly constant, always fluctuating around some number $P_0$, say $P_0 =
1000$, by applying the following death rule after the whole population
breeding is over. Each individual $i$ survives with probability $p =
x^{N_i+1}$, where $N_i$ is the number of homozygous 1-bit pairs in its
genome (1-bits in the same position of both $B$-strings). Before to kill
anybody, the value of $x$ is determined in order to give an overall death
rate compensating the appearance of newborns, i.e. by solving equation

$$x\,\sum_{i=1}^P x^{N_i} = P_0\,\,\,\, ,\eqno(2)$$

\noindent where the sum runs over all individuals (including newborns),
or alternatively

$$x\,\sum_{N=0}^\infty H(N) x^N = P_0\,\,\,\, ,\eqno(3)$$

\noindent where $H(N)$ counts the current number of individuals with $N$
homozygous 1-bit pairs. Once the value of $x$ is already known, the death
roulette is applied sequentially to each individual $i$, according to its
$N_i$. We count one more time step (one generation) after this complete
process, breeding and deaths, is applied to the whole population.

        This simple rule strictly follows Darwin's principles [2]. First,
the offspring genomes are slightly modified copies of the parents', with
the additional crossing mechanism observed in diploid real organisms.
These minor modifications from parent to offspring are performed at
random. No further genome modification is performed during each
individual's lifetime (no somatic mutations are considered). Second, again
following Darwin, selection acts through the different survival
probabilities, depending on some characteristics of the individual's
genome. Here, specifically, the larger the number of homozygous 1-bits it
has, the lower is its life expectancy, and thus the lower is the number of
offspring it is supposed to breed. In this case, 1-bits represent harmful
mutations. Many different versions of such a rule can be invented, the
first one being the famous Eigen model [3]. The details of each model are
not relevant here, the important feature is their common purpose to mimic
Darwin's evolution working in real time. A few among them allow analytic
treatments by translating the rules into differential equations, and
solving them.

        These analytic successes are rare. For most models, like the
present one, analytic approaches are hard or impossible: in spite of its
simplicity, one cannot translate the above rules into differential
equations. Then, another language is necessary. For instance, C, FORTRAN,
PASCAL or any other computer language which allows one to program the
machine to follow strictly the dynamic rules, starting from some given
population, during many time steps. By using some tricks [4], one can
repeat this routine many times with different randomnesses, in a very fast
way. Thus, one can measure the various properties of interest, their
averages and correlations, how much they fluctuate due to randomness, etc.
More important is the possibility to follow the historical evolution, time
after time, an impossible task concerning most examples of real biological
evolution. Moreover, one can even {\bf repeat} the historical path
introducing some controlled modifications, in order to study their
influences. We were able to follow the above defined model for $S = 1000$
different samples, with $P_0 = 1000$ individuals each, with a length
genome of $B = 1024$, during $t = 8192$ time steps (generations). First,
we have performed some preliminary computer runs starting from a
population with only 0-bits (no harmful mutations), storing at the end the
corresponding populations in order to use them later, as starting points.
We observed the same probability distribution of 0 and 1-bits, for all
these final populations, after $t = 8192$ generations, in spite of the
different randomnesses during their past evolutions. This statistical
equilibration is always achieved, provided the number of generations is a
few times larger than $P_0$, even starting from different initial
populations (with randomly assigned harmful mutations, instead of only
0-bits).

        Now, re-setting $t = 0$ for the already equilibrated populations
obtained from the preliminary runs, we performed the $S = 1000$
observation runs. A different family name is given to each individual at
generation $t = 0$. Different randomnesses were used during the time
evolution of each sample. For all of them, we observed that all final
individuals at generation $t = 8192$ share the same family name (inherited
without mutations from the mother). In other words, after $t = 8192$
generations, all the $P \approx 1000$ alive individuals are relatives of
each other, all descendents from the same original
grand-grand-$\dots$mother. Let's call this lineage founder individual,
which lived at generation $t = 0$, the Eve. Each sample has its own Eve,
different from sample to sample. Even by starting just from the same
initial population, different randomnesses lead to different Eves. This is
not a novelty, it is in complete agreement with the coalescence theory, a
mathematically well established set of branching-processes theorems
concerning genetic behaviour (see, for instance [5], for a soft an
excellent review). Other similar models also present this behaviour [6,7],
as expected. Some results of our computer simulations are in figure 1a.
The open circles show how the number of not-yet extinct family names
evolves in time, averaged for all $S = 1000$ samples. This number
monotonically decreases from the initial value ($\approx 1000$) at $t =
0$, reaches the value 1 after $\approx 1000$ generations, and remains
constant thereafter. The values on the vertical axis, figure 1a, are
normalised (divided) by the initial population. Also, only the first 64
generations are shown, but the monotonic decay goes on.

        The obvious question concerns the population genetic diversity. Is
it compromised by the fact that all final individuals are relatives,
belonging to the same family? Is the final population genetic pool
dominated by Eve's genome in some sense? In order to answer these
questions, we have measured the correlation between each individual's
genome and that of Eve, during all the time. In order to perform this
task, we have profited from the possibility of repeating the whole
evolution again, starting from exactly the same initial population and
following exactly the same randomness, after knowing who was Eve from a
first run. Only computer modeling allows such a procedure. We simply count
the number of coincident bits at the same position in both genomes of Eve
and the individual $i$. We performed this counting twice, first comparing
the first (second) $B$-string of Eve with the first (second) $B$-string of
$i$. Then, we repeated the counting by comparing the first $B$-string of
Eve with the second $B$-string of $i$, and vice-versa. The largest between
these two counters is normalised (divided) by $B$. The result (a number in
between 0 and 2) is the genome correlation. It measures how much the
genome of $i$ is ``similar'' to Eve's. At each generation, we measured the
average of this quantity for all alive individuals, and subtracted this
from the restricted average performed only among Eve's descendents. The
absolute value of this quantity is shown by pluses in figure 1a. It decays
to zero much faster than the number of family names. This means that,
within Eve's family, the same genetic diversity of the whole population is
reached in few generations, far before it becomes the only survival
family. Nature is smart enough to preserve (and fast restore, if
necessary) genetic diversity, even being forced to follow the coalescence
theorems which dictate that all individuals belong to the same family, at
the end.

        We invented the above defined (toy) model only in order to analyse
the deep mathematical differences between the two decaying curves in
figure 1a. The reader can forget the model hereafter. Later, we will
return to the much simpler time-evolution defined by equation (1), which
nevertheless presents all the essential features one can study about
dynamic systems, at least within the purpose of the present work. This
purpose is to discuss how Nature knows to deal with the two possibilities
exemplified in figure 1a, profiting from one or other mathematical
behaviour, in what seems to be a universal law governing all evolutionary
systems. This curious mathematical issue attracts the attention of
researchers at least since [8]. The text is divided as follows. In section
II, we analyse the mathematical properties of both ``experimental'' curves
shown in figure 1a, and some implications in what concerns evolutionary
behaviour. In section III, we return to equation (1) which also presents
both mathematical behaviours, depending on the value of the parameter $a$
fixed by the user. This equation has the advantages of allowing some
analytical treatment and of being much more easily programmed on a
computer. In section IV, we present our main argument. Conclusions are in
section V.

\vskip 50pt
{\bf \item{II)} Fast $\times$ Slow Decays, or\par
Exponential $\times$ Power Law Behaviours}\parG

        Figure 1a shows two decaying quantities, one fast (pluses), and
the other slow (circles). They are different averaged quantities
concerning the same dynamic population model, and were measured from a
series of computer simulations. Pluses display the genetic similarity
between a founder individual (Eve) which lived at generation $t = 0$ and
all its descendents alive at generation $t$. Let's call $g$ this quantity.
Its fast decay gives rise to a fast spread of genetic diversity among the
whole population. Eve's descendents fast ``forget'' their genetic origin:
the in-family differences between their genomes fast become statistically
the same as the whole population. In order to infer the mathematical form
of this ``experimental'' curve, we adopted the trick of showing the same
data in figure 1b, now plotted according to a logarithmic vertical scale
(equally spaced powers of 10). The first obvious effect of this trick is
to enhance for the eyes the $g$-fluctuations appearing after generation $t
\approx 20$, hardly observable in figure 1a. These fluctuations will
remain forever, and are due to the simple fact that both the population
size ($P_0 = 1000$) and the number of statistically distinct samples ($S =
1000$) are finite. Indeed, from the product $P_0S = 10^6$, one can expect
fluctuations around $1/\sqrt{P_0S} = 10^{-3}$, as really observed in
figure 1b. These fluctuations are not important at all in what concerns
our conceptual interpretation: would we have simulated a larger population
size with a larger number of samples, they would be pushed to below any
previously defined tolerance one wants.

        The really important feature of adopting a non-linear, logarithmic
vertical scale in figure 1b is the straight line observed before the
fluctuations dominate the scenario. From it, we can infer that $g$ follows
an exponential decay, i.e.

$$ g_t = g_0\, {\rm e}^{-\lambda t}\,\,\,\, ,\eqno(4)$$

\noindent where ${\rm e} = 2.718282\dots$ is the natural basis for
logarithms, and the constant $g_0 = 1$ is unimportant. From the slope of
this plot, fitted between $t = 0$ and $t = 20$ (dotted line), we can also
obtain the ``experimental'' value for the constant $\lambda$ which depends
on the mutation rate $M$. In this case, we found $\lambda \cong 0.21$.

        Circles display the fraction of yet-alive families (lineages)
relative to the founder population at $t = 0$ (each individual at $t = 0$
is a family founder, and pass its family name to its descendents). Let's
call $f$ this fraction. It decays much slower than $g$, and certainly does
not fit to an exponential function. The mere $64$ generations shown along
the horizontal axis common to both figures 1a and 1b are not enough to
fully observe the decay of $f$. Considering the observed fact that only
one family eventually survives, among the initial number $P_0 = 1000$, we
can conclude that the current value of $f$, at $t = 64$, is still far
above the final, definitive figure $10^{-3}$. Thus, in order to better
observe the slow decay of $f$, let's adopt the trick once more, now also
along the horizontal axis. Figure 1c shows the same data again, with
logarithm scale in both axis. The difference is the much larger number of
generations shown now, in figure 1c, up to $t = 8192$. This is a too large
number of points for a plot, thus we show only a few of them, namely those
corresponding to generations which are integer powers of 2 (t = 1, 2, 4, 8
$\dots$ 8192), for clarity.

        Now, in figure 1c, the fraction $f$ of alive families fits to a
straight (continuous) line after some few initial generations, leading to
the mathematical relation

$$ f_t = f_1\, t^{-\alpha}\,\,\,\, ,\eqno(5)$$

\noindent called a power law, where $f_1$ is another unimportant constant.
What matters is the universal value $\alpha = 1$, predicted by coalescence
theory for a very wide class of systems to which our toy model belongs.
Measuring the slope of our ``experimental'' curve, at its inflection
point, we get $\alpha \cong 0.98$ in agreement with the theory. One can
modify the particular, naive dynamic rules described in last section for
our toy model, as one wants, and the universal value $\alpha = 1$ will
insistently re-appear. More important, it also describes many actually
observed real population data. It is our first example of universality,
thanks to which some naive models can describe real situations. We will
return to this fundamental point, later.

        The important concept to be taken into account is the deep
difference concerning the fast exponential behaviour ${\rm e}^{-\lambda
t}$ {\it versus} the slow power law $t^{-\alpha}$. It is so deep, as we
will see later, that it is rather a qualitative distinction, not merely a
quantitative difference. It concerns the distinction between any time
interval, no matter how long it is, and eternity. Any evolutionary process
must be eternal (if it stops after some finite time, it no longer
evolves). Thus, evolution is expected to be mathematically described by
some power laws. It is interesting to note that, besides power laws,
Nature also uses the fast exponential behaviours, when necessary within
each particular evolutionary system (for instance, in order to restore
genetic diversity in our model). However, the quantities related with
these exponential behaviours are not universal, as the value $\lambda
\cong 0.21$, equation (4). It depends on the mutation rate $M$: the larger
$M$, the larger is $\lambda$ and faster Eve's genome is ``forgotten''.
These quantities also depend on the particular rule one adopts. For
instance, would we have considered haploid, asexual populations, then the
value of $\lambda$ would be smaller: driven only by mutations, without
recombination (crossing) within two distinct parents' genomes, the genetic
diversity spreads slower, although yet exponentially fast. On the other
hand, the exponents defining the power laws, as $\alpha = 1$ in equation
(5), are insensitive to particular parameters and universally valid for
wide classes of different systems and models.

        Also important is to note that a power law, equation (5) for
instance, never appears alone. When a given system presents such a
mathematical dependence between two particular variables, many other
quantities also depend on each other by power law relations with universal
exponents. In our model simulations, we have also measured the size of
each family, i.e. the total number of individuals which belonged to this
family during all its history. Then, we classified these families
according to their sizes, counting how many of them presented just 1
individual (no descendents at all, the smallest conceivable size), how
many presented 2 or 3 individuals (second smallest size), how many
presented 4, 5, 6 or 7 individuals, and so on. We choose family size
intervals increasing as the integer powers of 2, a traditional procedure
usually adopted in order to improve statistics. The occurrence of families
according to their sizes is plotted in figure 2, again a power law. A
similar counting was performed half a century ago by Gutenberg and Richter
[9], in what concerns the occurrence of earthquakes according to their
intensities. The number of already extinct biological genera as a function
of their lifetime [10] is another example. Many others can be found in
[11].

        Denoting by $s$ the family size and by $n$ their counter per size,
our ``experimental'' data in figure 2 leads to

$$ n = n_1\, s^{-\beta}\,\,\,\, ,\eqno(6)$$

\noindent where the new exponent $\beta = 1/2$ can be related to the
previous one, $\alpha = 1$, by following the same coalescence theory. By
fitting our data (dotted line in figure 2), we find $\beta \cong 0.50$
again in agreement with the theory. Normally, the various exponents
$\alpha$, $\beta$, etc, governing the universal behaviour of a whole class
of systems are not independent, but related to each other by some simple
relations.

        One can argue that, in practical terms, the power law behaviour in
equation (5) cannot continue eternally, once the number of alive families
eventually drops to a single, definitive one. However, this is only due to
the finite population size $P_0 = 1000$ of our example. Would we have
considered a 10-fold larger population, and the waiting time for
coalescence to a single family would also be multiplied by 10.
Accordingly, 10-fold larger family sizes would appear, and the plot of
figure 2 (or 1c) would suffer a translation to the right by adding one
more 0-digit to each horizontal axis label. A further equivalent
translation would occur for a 100-fold larger population, and so on. For
an ``ideal'' infinite population, coalescence will never occur, and all
family sizes and lifetimes would be expected to appear. In other words,
the absence of a size scale gives rise to the corresponding absence of a
time scale, and vice-versa. Other quantities also become scaleless. On the
other hand, the finite-size of a given system necessarily limits also its
characteristic scales of time, energy, mass, etc.

\vskip 50pt
{\bf \item{III)} Lessons from a Simple Example}\parG

        Let's return to the logistic map, equation (1), restricting
ourselves to the intervals $0 < x_0 < 1$ for the initial seed, and $0 < a
< 4$ for the control parameter. This last restriction forces the dynamic
variable $x_t$ to stay forever within the same interval $0 < x_t < 1$ as
the initial seed, in agreement with the population problem for which this
map was invented [1]. The reader can easily program equation (1) on
her/his pocket calculator, and appreciate the many different dynamic
behaviours one can get for the variable $x_t$, by choosing distinct values
for the fixed control parameter $a$. Indeed, this is just what was done
two decades ago by the now-famous physicist M.J. Feingenbaum.
Experimenting with equation (1) on his pocket calculator, Feigenbaum
discovered the so-called period-doubling route to chaos and the universal
behaviour valid for a huge class of distinct dynamic systems [12]. This is
another example of universality: precisely the same universal values
obtained by Feigenbaum from equation (1) were later measured in a lot of
real, different systems, as well as a lot of other, more complicated
computer models. Reference [13] is a friendly reading about this
discovery. Here, we will use the logistic map as a guiding example. One
can follow all our reasonings through numerical tests performed on a
pocket calculator or computer (indeed, I strongly recommend the reader to
do so). The concepts and conclusions, however, are completely general,
independent of equation (1).

        A first observation is that the (normalised) population $x_t$
vanishes for large values of $t$, when one chooses $a < 1$. On the other
hand, $x_t$ reaches a stable long-term value if $1 < a < 3$. There is a
transition between population stability and extinction, by crossing the
critical point $a = a_0 = 1$. In both cases, stability or extinction, the
attractor (i.e. the final destiny) is a single fixed point: $x^* = 0$ for
$a < 1$; or $x^* = 1 - 1/a$ for $1 < a < 3$. The transient time roughly
corresponds to the number of iterations one needs to perform from the
initial seed $x_0$ until reaching the final fixed point $x^*$, within the
machine accuracy. The closer the control parameter $a$ is to the critical
value $a_0 = 1$, the larger is this transient time. The particular
behaviour {\bf at} the critical point will be discussed later. For the
moment, let's choose $a$ {\bf near} 1. One can verify that $x_t$ evolves
according to an exponential decaying function

$$x_t - x^* \sim {\rm e}^{-|a-1|t}\,\,\,\, ,\eqno(7)$$

\noindent where $\sim$ means proportionality, and $|\dots|$ represents the
absolute value. This form can be obtained numerically, for instance by
plotting the logarithm of $x_t - x^*$ {\it versus} $t$, verifying that
this corresponds to a straight line (for large values of $t$), then
measuring its slope and comparing the result with $|a-1|$. We have already
followed this numerical recipe in figure 1b and the corresponding equation
(4). In the present case, alternatively, one can rewrite equation (1) as

$${{\rm d}y\over{\rm d}t} = - |a-1| y - a y^2\,\,\,\, ,\eqno(8)$$

\noindent by introducing a new variable $y_t = x_t - x^*$, where the
difference $y_{t+1} - y_t$ was identified with the derivative ${\rm
d}y/{\rm d}t$ (this procedure is valid for large values of $t$). Provided
$a \ne 1$, i.e. the system is {\bf near} but {\bf not at} the critical
point, one can neglect the quadratic term $a y^2$ in (8), compared with
the other, larger term $|a-1| y$. Then, the solution of this simple
differential equation is just expression (7).

        The exponential decay (7) allows us to define precisely the
transient time

$$\tau = {1\over|a-1|} = |a-1|^{-1}\,\,\,\, ,\eqno(9)$$

\noindent as the time during which the difference $x_t - x^*$ decays by a
factor of ${\rm e} \approx 2.7$. Alternatively, we can resort to an
analogy with a radioactive sample, for which $\tau$ corresponds also to
the mean lifetime

$$\tau = {\int_0^\infty {\rm d}t\,\, t\, {\rm e}^{-|a-1| t}
\over\int_0^\infty {\rm d}t\, {\rm e}^{-|a-1| t}} = {1\over|a-1|}\,\,\,\,
,\eqno(10)$$

\noindent i.e. the time one needs to wait for the radioactive emission
from a particular nucleus to occur, on average. In practice, $\tau$ is
roughly the time one needs to wait for the radioactive sample become
dangerless. In general, any system obeying an exponential decay has a
characteristic time $\tau$ well defined by equation (10), representing its
natural time scale, during which all important phenomena occur. One does
not need to consider times much larger than $\tau$, because they have no
effect on the system behaviour. In other words, $\tau$ roughly measures
the system's lifetime, after which all activities cease.

        Nature was kind enough to choose the exponential mathematical form
for radioactive decays: because of that, the integrals in equation (10)
converge, and we get a {\bf finite} value for $\tau$ (maybe large, but
finite). This is not the case for other dynamic systems at critical
situations, for instance the logistic map with $a = a_0 = 1$. In this
case, one can no longer neglect the quadratic term $y^2$ in equation (8),
because the other term $|a-1| y$ vanishes, and the solution of this
differential equation is now the power law decay

$$x_t - x^* \sim t^{-1}\,\,\,\, ,\eqno(11)$$

\noindent instead of the exponential form (7) valid for non-critical
situations. Trying to replace the exponential form appearing twice in
equation (10) by the power law (11), one would be in trouble because the
integrals no longer converge to finite values. Accordingly, one can no
longer perform the division in equation (9): the transient time is now
{\bf infinite}. These critical dynamic systems do not have a
characteristic time: all time scales are important. Nature is not always
so kind as in the case of radioactivity. For instance, the probability of
having an earthquake decays for increasing intensities, obeying a power
law (in this case, earthquake intensity replaces the time). This means
that there is not a characteristic earthquake intensity beyond which the
probability of occurrence can be neglected. All intensities are expected
to occur some day. Would this distribution be an exponential, engineers
could design buildings strong enough to support the characteristic
intensity, and earthquakes would be not such a big problem as they are.
Unfortunately, this characteristic earthquake intensity does not exist,
according to the power law behaviour reported by Gutenberg and Richter
[9]. Again, due to the fact that the Earth itself is finite, some
earthquake intensity upper bound certainly exists, beyond which the
Gutenberg and Richter law is no longer valid. However, nobody knows what
could be this finite-size characteristic intensity: large, extremely
destructive earthquakes could occur still below this unknown bound.

        The distinguishing feature of the isolated critical point $a = a_0
= 1$, among all continuous distributed possibilities $0 < a < 3$, is the
{\bf eternal transient} followed by the dynamic system, equation (1).
Mathematically, instead of an ordinary exponential decay which limits the
system history to a finite lifetime, at critical situations one has power
law decays without time limits. This mathematical feature is completely
general for all dynamic systems, and is indeed taken as the definition of
criticality. At these situations, the system presents {\bf long-term
memory}, i.e. its current state is a consequence of many features
accumulated during a long past history. Many ``strange'' properties appear
{\bf only at} these situations. For instance, a widespread class of
numerical devices called relaxation methods consist in finding the true
solution of some problem by gradually performing small modifications (or
mutations) on an initially posed approximation. It is a dynamic path to the
solution. In principle, it would be enough to design a dynamic rule whose
attractor has been previously shown to coincide with the desired solution.
However, if the chosen dynamic rule is critical, the user will be in
trouble. First, because the critical slowing down forces the computer time
to be prohibitively large. Second, and much worse, because the final
(finite precision) numeric answer will be wrong! The reader can verify
both characteristics within the logistic map with $a = a_0 = 1$. In
particular, in case the reader is patient enough to wait for convergence,
the final reached value $x_\infty$ will differ from the right answer $x^*
= 0$ by half the digit precision of her/his pocket calculator --- half of
the digits will be wrong because of the square in equation (8). Also, the
wrong digits depend on the initial seed introduced at the very beginning,
a symptom of the long-term memory.

        The concepts of short {\it versus} long-term memory can be better
understood by performing a simple exercise on both equations (4) and (5):
try to express the next value $g_{t+1}$ (or $f_{t+1}$) as a function of
only the current one, $g_t$ (or $f_t$). In the exponential case (4), this
is possible, and the result is simply $g_{t+1} = {\rm e}^{-\lambda} g_t$.
This means that $g_0$, $g_1$, $g_2$, $\dots$ is a Markovian sequence, i.e.
each term depends only on the previous one, not on remote past terms like
$g_0$. The system ``remembers'' only one past step. In the power law case
(5), however, the exercise is fruitless: the best one can do is to write
$f_{t+1} = [f_t^{-1/\alpha} + f_1^{-1/\alpha}]^{-\alpha}$, i.e. the next
term $f_{t+1}$ depends not only on the previous one, $f_t$, but also on
the very first, $f_1$. The system always ``remembers'' its birth, no
matter how old it is. Note, however, that both $g$ and $f$ are quantities
concerning the same dynamic system, namely our toy population model
treated before. Note also that the system itself {\bf is} Markovian,
because the next generation is constructed exclusively from the current
one. Thus, even a Markovian system can exhibit long-term memory behaviour
concerning some particular quantities which follow power law decays. Other
exponentially decaying quantities could also coexist within the same
system, for which the memory is instead short term. Of course, this
simultaneous occurrence of both forms is not possible for simple systems
as the logistic map, which presents only one varying quantity, $x_t$: for
some values of the fixed parameter $a$ its memory is short, as in equation
(7), for others it is long-term, as in (11).

        Critical, long-term memory situations normally occur at precisely
tuned transition points where the system behaviour changes qualitatively,
such as the temperature above which a magnet can no longer retain its
spontaneous magnetisation, or a fluid can no longer be found in both
(coexisting) liquid and vapour forms. These examples are static, in the
sense that thermal equilibrium is supposed to be already reached for both
the magnet and the fluid: universal power laws in space (instead of time)
govern the system behaviour. Of course, also the path previously followed
towards thermal equilibrium was governed by a universal power law decay in
time. As before, the term ``universal'' means the same numerical values
for some characteristic exponents, like the $-1$ in equation (11), holding
for completely distinct systems. Both the fluid compressibility and the
magnetic susceptibility diverge at the critical point according to the
same critical exponent, valid for a huge class of fluids and magnets. This
universality is a consequence of the lack of both a characteristic time
and length scales: most microscopic or short term details are not
important for those scaleless (critical) systems. Often, a very simplified
model is enough to reproduce, even in a quantitative sense, the complex
behaviour observed in a much more complicated, real system. For instance,
as we have already treated, simplified mathematical models based on
Darwin's principles could allow one to follow evolution features step by
step, in real time --- see [14] for a review.

        The theoretical explanation for the universality observed in
equilibrium critical phenomena is due to the Nobel laureate K.G. Wilson
[15]. Concerning critical dynamics, an equivalent complete theory is still
lacking. Nevertheless, it is well known that systems evolving according to
a critical dynamics (i.e. a time decay to equilibrium according to a power
law) normally also present static critical behaviour (i.e. power law
relations between quantities other than time). Due to the long times
involved in such situations, the dynamic critical behaviour of many
systems can only be observed indirectly, through these other timeless
power law relations. Biological evolution is an example, where the already
quoted statistics of mass extinction are found to obey power laws, both
from fossil records data [10] and mathematical models [16]. Earthquakes
intensity distribution is another example. In most cases the dynamical
criticality is the fundamental one, although hard or impossible to follow
step by step. The long-term memory of the system is responsible for its
whole critical behaviour. Although very interesting and important, the
universal features found in critical systems are not directly linked to
our main argument concerning evolution strategies. Before entering into
this point, let's stress that many dynamic systems are able to adapt
themselves according to the environment, by self organizing some of their
own control parameters, with no need to resort to external tuning
mechanisms. In particular, a very common feature is the so-called self
organized criticality, where the system keeps itself at critical
situations [17]. Among others, biological evolution was proposed to belong
to this class of system [18] (see also [19] for an excellent explanation).

        Back to the logistic map, there are other critical points. The
next one is $a = a_1 = 3$, beyond which the attractor is no longer a
single fixed point. Instead, for $3 < a < 1+\sqrt{6}$ the attractor is a
cycle with period 2, i.e. a sequence of two alternating values $x_1^*$,
$x_2^*$, $x_1^*$, $x_2^*$, $x_1^*$, $x_2^*$ $\dots$. The next critical
point is $a = a_2 = 1+\sqrt{6} \approx 3.449$, after which the attractor
become a 4-period cycle. There is a cascade of critical points $a_0$,
$a_1$, $a_2$, $a_3$ $\dots$, where successive period doublings occur. This
cascade ends at $a_\infty \approx 3.570$, after which one can find chaotic
behaviour, with no longer periodic repetitions. Even inside the chaotic
region $a_\infty < a < 4$ some periodic ``windows'' appear again, as one
just after $a = A_3 = 1+\sqrt{8} \approx 3.828$ where a period-3 cycle
holds. Thus, one has a series of transitions from one kind of attractor to
another (vanishing population, fixed point, periodic cycles, chaotic
attractors, odd-periodic cycles, chaotic again, and so on). The system is
critical only at these transition points, but still presents exponential
decays (and thus finite transients) in between two such transition points,
including the chaotic intervals. The chaotic behaviour concerns the
diverging character of two initially neighbouring trajectories, having
nothing to do with how fast both reach the attractor. In what concerns the
transient lifetime, both regular (fixed points or periodic cycles) and
chaotic systems are not critical, both approaching the corresponding
attractor exponentially fast, according to short term memories. So, we are
not interested in classifying dynamic systems in chaotic or regular, but
in distinguishing the very particular critical, long-term memory
situations among the whole set of possibilities. For the logistic map, for
instance, these situations occur only at the very precise values $a_0$,
$a_1$, $a_2$, $a_3$ $\dots$ $a_\infty$ $\dots$ $A_3$ $\dots$ of the
control parameter. Figure 3 shows the so-called Lyapunov exponent
measuring the exponential rate of convergence (negative values) or
divergence (positive) of two initially slightly different trajectories, as
a function of the control parameter $a$. The critical situations
correspond to zero Lyapunov exponent, at the particular positions where
the plot crosses (or touches) the horizontal axis. At these situations,
the rates of convergence or divergence no longer follow the fast
exponential way. Instead, it is replaced by the power law way, and further
mathematical analysis [20] is needed. This analysis is similar to what we
have done by keeping explicitly the term $-y^2$ in equation (8) for the
particular case $a = 1$, obtaining the power law (11). For most values of
$a$, however, the system is either regular (below the horizontal axis, in
figure 3) or chaotic (above), both cases corresponding to non critical,
short term memory behaviour.

\vskip 50pt
{\bf \item{IV)} Why do Evolutionary Systems Stick to the Edge of
Chaos}\parG

        Evolution is an eternal, endless process. Thus, it cannot occur in
a closed, finite system because size limits correspond also to time
limits, according to our previous analysis. Any dynamic activity occurring
within a finite system completely isolated from the rest of the universe
will eventually cease, after a finite time. This ephemeral activity is
rather a transient before reaching final static equilibrium, not
evolution. A single species never evolves in isolation, it is always under
the influence of other species. Nevertheless, it is hard to conceive the
whole universe as the evolving system to be studied. The practical
approach is to follow only the behaviour of a small part of the universe.
Then, one needs to consider also the environment in which this finite
system is embedded. Such a system is finite, but not closed. Instead, it
is an {\bf open dynamic system}, constantly fed from outside with mass,
energy, heat, information, etc. It also constantly throws away these same
entities, after processing them in some way. It is a {\bf dissipative}
system. One needs to have always in mind that this separation of a small
part of the universe is only a working procedure, an artificial
approximation which could be very good up to certain limit scales of time,
size, energy, etc. Beyond these limits, this approximation fails, and some
parts of the former ``environment'' must be included into a new, enlarged
``system'' to be studied.

        Being dissipative, this kind of dynamic evolution tends to
converge to some attractor, becoming irreversibly trapped there forever.
One of the dissipated quantities during this process is the entropy, which
measures the number of available states, related to the current options
the system could follow from now on. Considering the whole initial space
of states, only a shrinking subset of them remain available, as time goes
by. The system continuously looses its ability to explore the whole space
of states. Eventually reaching the attractor, it becomes restricted to a
subspace of the whole initial set of possibilities. Compared with the
whole, this subspace is a null measure set, like a needle in a haystack:
by tossing a random state among all the initially available options, the
probability to become inside the final subspace is null. In other words,
the dimension of the final subspace of states is smaller than the
dimension of the initially available space, like a line inside a plane, or
a plane inside the 3-dimensional space. Note that such a dimensional
reduction is much drastic than a finite-fraction decrement, because what
remains at the end is a {\bf null fraction} of the whole. The final
attractor may also be a fractal, i.e. it may have a non-integer dimension,
anyway smaller than that of the whole space.

        Within the mathematical description of evolutionary systems, the
dimension of the space of states is normally very high, as we have
exemplified in section I. Even the simplest approach of counting the
frequencies of all possible alleles for each gene, storing the results on
an array, would correspond to a high dimension, namely the total number of
possible alleles. In order to study also correlations between these
alleles, a larger yet dimension arises. By storing separate informations
concerning each individual, as in our population dynamic model discussed
before, the dimension increases even more. As time goes by, the mere
extinction of a single allele would correspond to a dimension reduction
with its huge and irreversible trapping effect. Concerning evolution
strategies, such a dissipative dynamics is not convenient. Indeed, after
being trapped into the tiny attractor, the probability of exploring the
remainder set of possibilities (searching for forms better than the
current one) is null. On the other hand, evolution within a limited size
system cannot be described by a non-dissipative, conservative dynamics
which theoretically could solve the problem by keeping it forever visiting
all possible options. This situation would correspond to a closed object,
isolated from the rest of the universe, not to a finite-but-open dynamic
system we need in order to take the environment into account.

        How Nature solves this puzzle? Let's consider a wonderful example,
extracted from [21], an amazing text which treats exactly the same subject
as the present work, namely the importance of diversity, from a completely
distinct (and much more interesting) point of view. A recessive genetic
disease affects homozygous individuals carrying twice the defective
allele. An example is phenylketonuria, for which the current frequency of
the defective allele $p$ in France is estimated to be $x = 0.95\%$. Let's
suppose that all homozygous $pp$ individuals die before reproduction, and
that heterozygous $Np$ individuals do not present any handicap at all ($N$
represents the normal allele). It is obvious that the frequency $x$ of
this harmful allele $p$ will decrease, eventually becoming extinct.
However, only after 6 generations, i.e. one and a half century from now,
the current french figure would drop to $x = 0.90\%$. Moreover, after 95
generations, i.e. more than twenty centuries from now, the defective
allele frequency would drop to $x = 0.50\%$, still a very small
``improvement''. This extremely slow time decay for the frequency of a
defective allele could seem paradoxical, if one adopts the restricted
reasoning of considering the evolution of this allele alone, separated
from the whole set of genes within each individual. Natural selection,
however, acts on each individual as a whole, according to its global
characteristics. Within this example, $pp$ individuals die but the much
more numerous $Np$ individuals remain untouched, avoiding the extinction
of allele $p$ (by delaying it {\it sine die}). Taking the opposed
reasoning of considering the survival of the whole species, one can argue
that instead of a paradox, the slow frequency decay of $p$ is just the
expected behaviour, because this allele could have some beneficial
function in some far future, under unpredictable modified environment
conditions. Indeed, some examples of imune behaviour against some diseases
are known to hold among individuals affected by other genetic diseases.
According to this reasoning, Nature adopts the precaution of preserving
these (nowadays) ``bad'' genes.

        The figures given by Albert Jacquard [21] concerning the frequency
$x$ of phenylketonuria allele $p$ came from the Hardy-Weinberg fundamental
law of genetics. He simply supposes random parent pairings, according to
the frequencies $(1-x_t)^2$ for $NN$ and $2 x_t (1-x_t)$ for $Np$
individuals, at the current generation $t$, composing from these pairings
the next generation. For instance, one can sum up the probabilities
corresponding to all $pp$-offspring possibilities, equating the result to
$x_{t+1}^2$, from which one can obtain the frequency

$$x_{t+1} = {x_t\over 1+x_t} \cong\, x_t\, (1-x_t) \eqno(12)$$

\noindent for the $p$ allele at the next generation $t+1$. The Lyapunov
exponent of this equation is exactly zero, i.e. it corresponds to a
critical dynamics. The approximation on the right-hand side of (12) is
valid for small values of $x$. It is amazing, but not a surprise, to
verify that it is just the logistic map (1) {\bf at its critical point} $a
= a_0 = 1$. Thus, Nature's strategy to avoid the extintion of $p$ consists
in adopting a critical dynamic evolution for it, according to which its
frequency power law decay is very slow, following an {\bf infinite}
transient time. Within the genetic high-dimensional space of the evolving
population, the particular direction corresponding to $p$ allele remains
always available, due to a small but never null occupation. The critical
dynamics avoids the dimensional reduction which would be caused by an
eventual $p$ extinction. Under a practical point of view, as the
population is finite, the extinction time for $p$ would be also finite,
nevertheless proportional to the average number of individuals, as we have
already analysed, and so very large.

        For evolutionary purposes, it is imperative to avoid the tiny
attractor provided by the dissipative dynamics, and this is an automatic
feature at the critical situations for which the system evolves forever in
transients. The evolutionary system naturally tune their internal
parameters in order to stay at such a critical situation. The tiny
attractor can be interpreted as the current ``best'' form, in singular.
However, the selection mechanism needs diversity in order to obtain the
current ``best'' forms, in plural, i.e. some enlarged neighbourhood around
the tiny attractor. Critical dynamics provides just this feature: one
stays forever {\bf near} the current ``best'', but never trapped into
their tiny cage. In this way, any environment modification which slightly
changes the ``best'' position of the space of possibilities, is likely to
match some other neighbouring form kept alive together the former
``best''. Any eugenic attempt to de-tune the evolving system out from its
critical situation, performed in order to accelerate the selection of the
``best'' form, again in singular, could be disatrous [22]. Indeed, this
procedure would replace a broad, critical distribution of many coexisting
forms by a sharp distribution of only one ``best'' form. The de-tuned
non-critical dynamics drives the system to this optimized, single form,
according to a fast exponential decay. This process could be considered an
efficient, short term optimization strategy, but it definitely does not
correspond to evolution. The selection mechanism cannot proceed from this
point on, due to the lack of diversity: the now-extinct neighbouring forms
could miss under some future environment change. Critical dynamics provide
the degree of diversity the selection process needs. If the dynamic
process is not critical, one has no diversity, thus no selection, and
finally no evolution.

        In short, evolutionary systems stick to critical situations (the
edge of chaos), because within all other possibilities (regular or chaotic
regimes) they rapidly become trapped into low-dimensional attractors,
loosing diversity. This trapping feature forbids the system to explore the
whole set of available possibilities inside its higher-dimensional space
of states. Although being of fundamental importance, this feature is only
the beginning of the whole story about evolutionary strategies. In order
to design an efficient dynamic evolutionary strategy, one must first be
sure to stay always at critical situations. But this is not enough! By
simply following a critical one-dimensional trajectory, one would not
explore the whole space of possibilities. It would be very hard to explore
a high-dimensional space by following one-dimensional trajectories:
infinitely many of them must be tested. In order to overcome this
difficulty, a lot of alternative optimization strategies exist: some
famous are simulated annealing [23], its generalizations [24] and genetic
algorithms [25]. Among these and many others, we will quote only the one
Nature has chosen for biological evolution, namely the endless
mutation/selection approach discovered by Darwin more than a century and
half ago [2] --- see also [26].

        According to Darwin, random mutations occur during reproduction,
the offspring being slightly different from their parents. Translated to
our language, and considering only asexual reproduction for simplicity,
this means a single point (the parent) in our space of possibilities
generating many other neighbouring points (the offspring), in a branching
process. Considering the offspring's offspring, and so on, this would
generate a ramified tree eventually covering the whole space of
possibilities {\bf in parallel}, optimizing the search for better forms.
However, this would cause population explosion, and some branches of this
tree are actually cut by natural selection, i.e. the individuals best
adapted to the current environment have more offspring (growing branches),
while the worst adapted ones tend to die without generating offspring
(dangling branches). Instead of a single one-dimensional trajectory, this
tree is a low-dimensional object avoiding populations explosion, the
growing branches exploring the most promising parts of the whole space of
possibilities. The dimension of this tree is itself a varying quantity,
allowing the occurrence of suddent large events. Of course, this evolution
strategy would not work at all if the dynamic would not be a critical one.
Thus, the natural selection process also acts in driving the dynamic
evolution to critical situations, avoiding trappings. This can be done in
many ways. Within an evolving population, for instance, the average birth
rate (equivalent to the logistic map parameter $a$) can be self-organized
into critical values.  Other parameters can also play the same role, and
after some external modification, they can be re-tuned according to the
new environment conditions. This automatic re-tuning mechanism towards
criticality introduces a new level into the whole dynamic process.
However, this re-tuning must be performed according to a fast rate, not to
be confused with the low rate observed when criticality is already
achieved. Once more, Nature is challenged to deal with both exponential
(fast) and power law (slow) decaying rates.

        As we have already discussed, the reasoning of last paragraph is
not complete. The particular evolving species we considered is not alone.
Think about the coevolution of many species. The space of possibilities   
where one particular species evolves is not a static landscape. This 
search is a branching process performed over an {\bf evolving landscape},
because the environment itself depends on the evolution of other species.
Thus, the search for better forms is an endless process: one can be
comfortably sit on an optimum state today, i.e. the best possible
situation among the whole neighbourhood, but tomorrow this same state may
be worse than some neighbouring competitor, due to the slow movement of  
the underlying landscape. Under the global point of view, the above quoted
re-tuning mechanism of parameters within each species is actually an
internal job belonging to the evolution of the global system. They are not
parameters controling the behaviour of each species separately, they are
dynamic variables internal to the whole set of coevolving species. Under a
larger yet point of view, a whole system of interacting species, for   
instance the ones living in Madagascar, is also influenced by other
similar systems. Thus the very particular set of ``parameters'' allowing  
for the odd wildlife observed in Madagascar is only, actually, part of the
current state of the flora and fauna evolving in larger Africa. Moreover,
the oddness of Africa itself is part of the current state of the evolving
Earth or Gaia, again increasing once more the time and size scales, and so
on (instead, we would have written earth and gaia). That is why evolution
occurs in a scaleless, critical way, both in time and space.

        Together with random mutations and natural selection, this
self-organized criticality in now considered as a third leg needed to keep
stable the basis for biological evolution theory, as reviewed in [19].
Although catastrophic events are crucial for the behaviour of critical
dynamic systems, they do not play any special role as compared to small
events (both follow the same scale-free probability distribution power
law). However, the study of this amazing subject goes beyond the scope of
the present work. Indeed, only recently it was recognized as an inportant
new scientific deep stuff, as beautifully presented in [11]. The important
role played by extremal dynamics is reviewed in [27].

        Nevertheless, this third leg is not a further ingredient, being a
consequence of the other two. Mutations and selection act on the
individual, microscopic level, following a gradual smooth rate. However,
it is not correct to conclude that their effect on the macroscopic level
of the whole population would be also gradual and smooth. This wrong
conclusion is based on the usual linear reasoning according to which the
sum of small increments will be also small. However, this reasoning does
not hold when one adds infinitely many small increments, whose sum may be
very large. This is just the case of long-term memory systems, which
``remember'' infinitely many small steps from the past. Evolution is not a
linear system. Indeed, this is an old puzzle since Darwin, due to the fact
that fossil data show some catastrophic biological extinctions as well as
periods of large speciation activity, the so-called punctuated equilibrium
[28]. Due to the long-term, critical memory, small modifications at the
microscopic level can accumulate themselves during long periods of time,
suddently overflowing into a fast catastrophic event. This feature,
however, is not due to a third, independent ingredient which would be
missing in original Darwin's principles. On the contrary, criticality is a
long-term consequence of the short-term mutation/selection process
repeated {\it ad infinitum}. Of course, the idea of catastrophic, fast
events being a consequence of infinitely many small steps, accumulated
during an infinite past, has emerged only after computers allow us to
follow evolution models step by step, in real time. Darwin could not
imagine such a non-linear mechanism at his time, not because it would not
be a consequence of his principles (it is!), but because he missed the
right instrument, i.e. the computer, to drive his imagination a little bit
further.

        A very common argument against evolution theory is based on some
estimates of the time needed to get the biological complexity observed
nowadays, comparing them with the age of the universe. The
counter-argument is that, in performing such estimates, one cannot take
into account all the set of potential possibilities, but only the ones
already selected by the critical branching process, at each step. As a
consequence, the estimated times are actually much smaller [19].

\vskip 50pt
{\bf \item{V)} Conclusions}\parG

        We have proposed here that evolutionary strategies must follow
critical dynamics in which slow power law decays replace the wide spread
fast exponential way to reach equilibrium. Equilibrium situations, even in
chaotic systems, represent a tiny fraction of the whole set of
possibilities. Thus, any system which rapidly reaches its equilibrium
looses the chance of searching for new, better forms in the
high-dimensional space of possibilities. On the other hand, by following a
critical dynamics, the system actually never reaches its equilibrium
situation, staying forever in transients. This opens for the system the
chance of eternal search for better forms, i.e. opens the whole space of
possibilities instead of only the tiny fraction represented by the
equilibrium states. This is a pre-condition for any efficient evolutionary
strategy, and can be fulfilled by self-organization of the parameters
controling the system's behaviour. The evolutionary strategy itself starts
at this point (the dynamic criticality already assured), and deals with
the problem of efficiently walking around a high-dimensional space through
low-dimensional trajectories. Many such strategies exist, including the
biological one discovered by Darwin, which naturally leads to a critical
dynamics. The important here is that none of these strategies would be
efficient if the corresponding dynamic rule would not be a critical one.

       Natural selection is based on the available diversity of options.
Both regular and chaotic dynamics drive the system quickly to its tiny  
attractor, where diversity is then irreversibly lost. Critical dynamics,
on the other hand, keep the system forever {\bf near} but {\bf never
restricted} to the tiny attractor, providing the necessary diversity which
allows the selection process to goes on, eternally. In practical terms,
for a finite system, ``eternity'' means some large time of the same order
of magnitude of the populations itself. Within a non-critical dynamics,
however, any activity ceases much before such a time, independently of the
population size.

\vfill\eject
\centerline{\bf Aknowledgements}\parG

        I am indebted to Suzana Moss, Jorge S\'a Martins, Kris Stenzel,
Am\'erico Bernardes, Ant\^onio Toledo Piza, Per Bak, Alaor Chaves, Jean
Lobry, Dietrich Stauffer and Moys\'es Nussenzveig for helpful discussions
and critical readings of the manuscript. This work was partially supported
by Brazilian agencies CAPES, CNPq and FAPERJ, and dedicated to the late
J.J. Giambiagi.

\vskip 50pt
\centerline{\bf References}\parG

\item{[1]} P.F. Verhulst, {\it M\'emoires de l'Acad\'emie Royale de
Belgique} {\bf 18}, 1 (1844).\par

\item{[2]} C. Darwin, {\sl On the Origin of Species by Means of Natural
Selection}, Murray, London (1859).\par

\item{[3]} M. Eigen, {\it Naturwissenschften} {\bf 58}, 465 (1971).\par

\item{[4]} P.M.C de Oliveira, {\sl Computing Boolean Statistical Models},
World Scientific, Singapore London New York, ISBN 981-02-0238-5
(1991).\par

\item{[5]} L. Excoffier, {\it La Recherche} {\bf 302}, 82 (1997).\par

\item{[6]} A.T. Bernardes and D. Stauffer, {\it Int. J. Mod. Phys.} {\bf
C6}, 789 (1995); P.M.C. de Oliveira, S. Moss de Oliveira and D. Stauffer,
{\it Theor. Biosci.} {\bf 116}, 3 (1997); F. Meisgen, {\it Int. J. Mod.
Phys.} {\bf C8}, 575 (1997); S. Moss de Oliveira, G.A. Medeiros, P.M.C. de
Oliveira and D. Stauffer, {\it Int. J. Mod. Phys.} {\bf C9}, 809
(1998).\par

\item{[7]} S. Moss de Oliveira, P.M.C. de Oliveira and D. Stauffer, {\sl
Evolution, Money, War and Computers: Non-Traditional Applications of
Computational Statistical Physics}, Teubner, Stuttgart Leipzig, ISBN
3-519-00279-5 (1999).\par

\item{[8]} B.J. West and M.F. Shleginser, {\it Int. J. Mod. Phys.} {\bf
B3}, 795 (1989).\par

\item{[9]} B. Gutenberg and C.F. Richter, {\it Ann. Geophys.} {\bf 9}, 1
(1956).\par

\item{[10]} D.M. Raup and J.J. Sepkoski Jr., {\it Science} {\bf 215}, 1501
(1982); D.M. Raup, {\it Science} {\bf 231}, 1528 (1986); R.V. Sol\'e, S.C.
Manrubia, M. Benton and P. Bak, {\it Nature} {\bf 388}, 764 (1997).\par

\item{[11]} P. Bak, {\sl How Nature Works: the Science of Self-Organized
Criticality}, Oxford University Press (1997).\par

\item{[12]} M.J. Feigenbaum, {\it J. Stat. Phys.} {\bf 19}, 25 (1978);
{\bf 21}, 669 (1979).\par

\item{[13]} M.J. Feigenbaum, {\it Los Alamos Science} {\bf 1}, 4
(1980).\par

\item{[14]} L. Peliti, {\sl Introduction to the Statistical Theory of
Darwinian Evolution}, lectures given at ICTP Summer College, Trieste, also
in Cond-Mat/9712027 (1997).\par

\item{[15]} K.G. Wilson and J. Kogut, {\it Phys. Rep.} {\bf 12C}, 75
(1974); K.G. Wilson, {\it Sci. Am.} {\bf 241}, 140 (August 1979).\par

\item{[16]} P. Bak and K. Sneppen, {\it Phys. Rev. Lett.} {\bf 71}, 4083
(1993); J. Maddox, {\it Nature} {\bf 371}, 197 (1994); R.V. Sol\'e and
S.C. Manrubia, {\it Phys. Rev.} {\bf E54}, R42 (1996); K. Christensen, R.
Donangelo, B. Koiller and K. Sneppen, {\it Phys. Rev. Lett.} {\bf 87},
2380 (1998).\par

\item{[17]} P. Bak, C. Tang and K. Wiesenfeld, {\it Phys. Rev. Lett.} {\bf
59}, 381 (1987); {\it Phys. Rev.} {\bf A38}, 364 (1988); P. Bak and K.
Chen, {\it Sci. Am.} {\bf 264}, 26 (January 1991).\par

\item{[18]} C. Langton, {\it Physica} {\bf D42}, 12 (1990).\par

\item{[19]} S.A. Kauffman, {\sl At Home in the Universe}, Oxford
University Press (1995); {\sl Origins of Order: Self-Organization and
Selection in Evolution}, Oxford University Press (1993).\par

\item{[20]} M.L. Lyra and C. Tsallis, {\it Phys. Rev. Lett.} {\bf 80}, 53
(1998).\par

\item{[21]} A. Jacquard, {\sl \'Eloge de la Diff\'erence: la
G\'en\'etique et les Hommes}, \'Editions du Seuil, Paris (1978).\par

\item{[22]} S. Cebrat and A. Pekalski, {\it Eur. Phys. J.} {\bf B11}, 687
(1999).\par

\item{[23]} S. Kikpatrick, C.D. Gelatt Jr. and M.P. Vecchi, {\it Science}
{\bf 220}, 671 (1983).\par

\item{[24]} T.J.P. Penna, {\it Phys. Rev.} {\bf E51}, R1 (1995); {\it
Comp. in Phys.} {\bf 9}, 341 (1995); D.A. Stariolo and C. Tsallis, in {\sl
Ann. Rev. Comp. Phys.} vol. II, ed. D. Stauffer, World Scientific,
Singapore London New York (1995).\par

\item{[25]} J. Holland, {\sl Adaptation in Natural and Artificial
Systems}, The University of Michigan Press, Ann Arbor, MI (1986).\par

\item{[26]} R. Dawkings, {\sl The Blind Watchmaker}, Norton, New York
(1986).\par

\item{[27]} P. Bak, M. Pakzuski and S. Malov, {\it Braz. J. Phys.} {\bf
24}, 915 (1994).\par

\item{[28]} S.J. Gould, {\sl The Wonderful Life}, Norton, New York
(1989).\par

\vfill\eject
\centerline{\bf Figure Captions}\parG

\item{fig. 1} Time decaying of two different quantities $f$ and $g$, both
concerning the same population dynamics. Each founder individual at
generation $t = 0$ inaugurates a family, and all its descendents carry its
family name: $f$ (circles) measures the still alive fraction of families.
After thousands of generations, only one family remains alive, all current
individuals being descendents from the same family founder, Eve. The
correlation (similarity) between Eve's genome and all other individuals
were also measured. Genetically, Eve is more similar to its own
descendents than the average among all individuals of the same generation.
The difference $g$ (pluses), however, decays with time much faster than
the number of families. Genetic diversity evolves very fast, even within a
very slow family number decay. 1a) linear scale in both axis; 1b) vertical
logarithm scale, same data; 1c) logarithm scale in both axis, same data
again, within a much larger time scale.\par

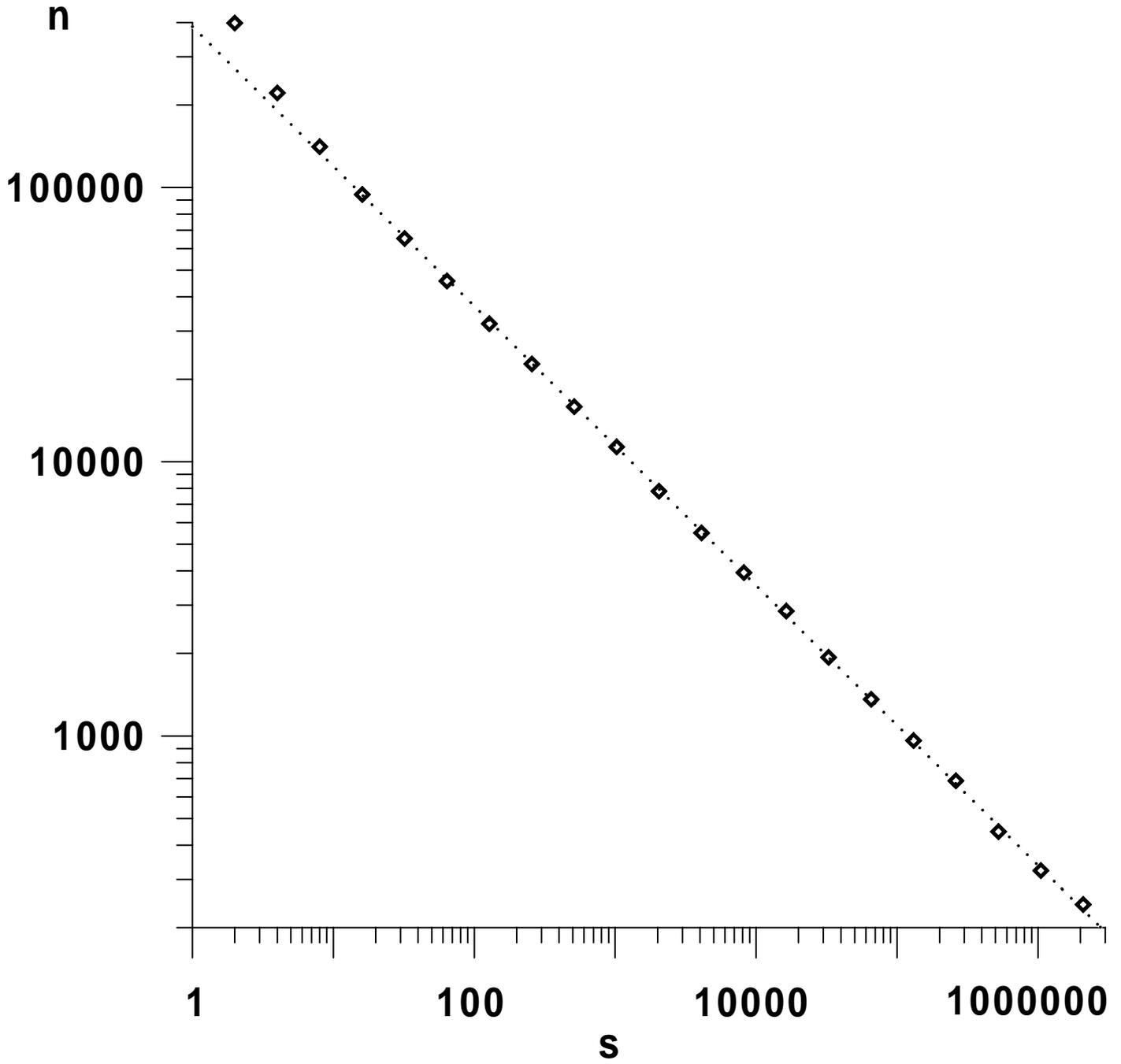
\item{fig. 2} For the same population dynamics corresponding to figure 1,
we measured the number $n$ of families classified according to their sizes
$s$ (total number of individuals sharing the same family name).\par

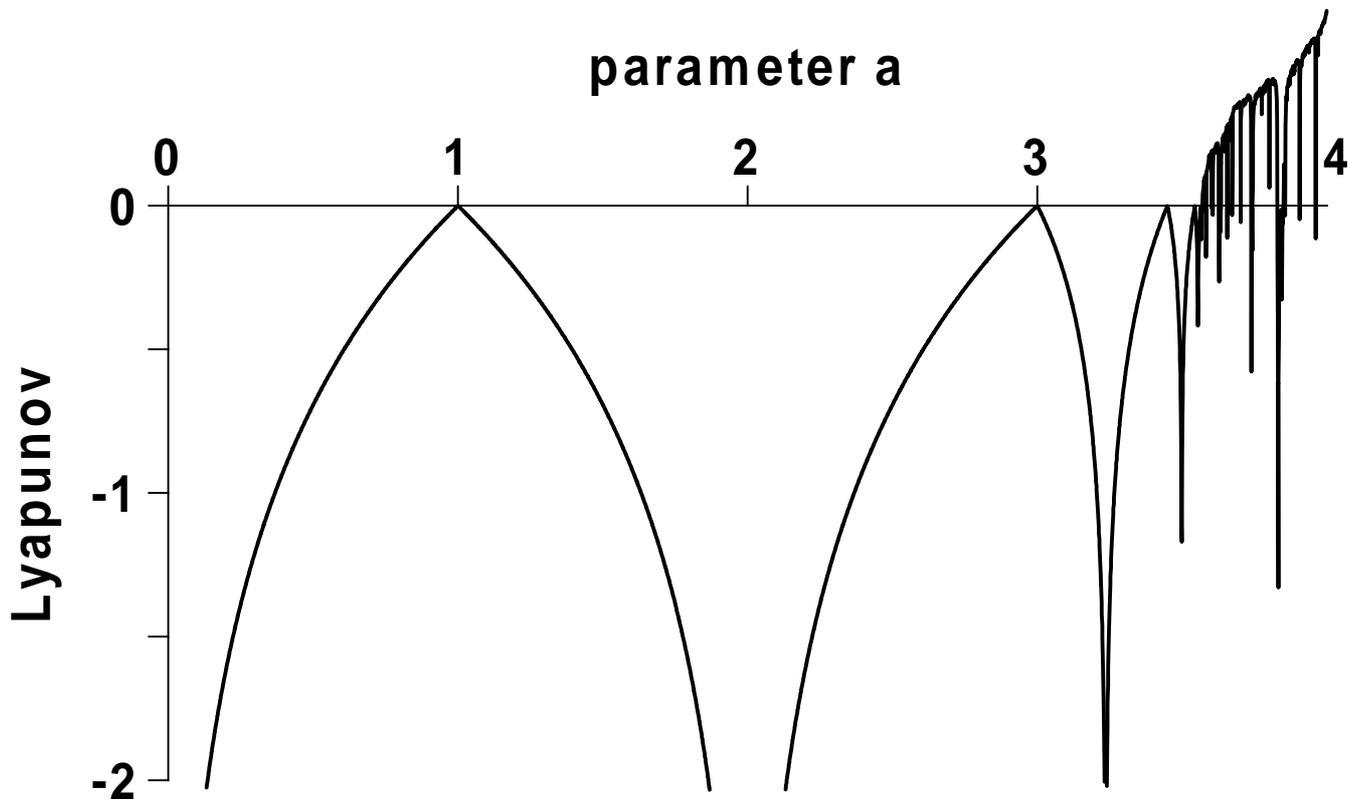
\item{fig. 3} Lyapunov exponent for the logistic map, equation (1), as a
function of the control parameter $a$. Below, a detail inside the interval
$3 < a < 4$. This dynamic system is regular below the horizontal axis
(negative exponents), and chaotic above (positive). Instead, the present
text concerns only the relatively few critical situations where the plot
crosses (or touches) the horizontal axis.\par

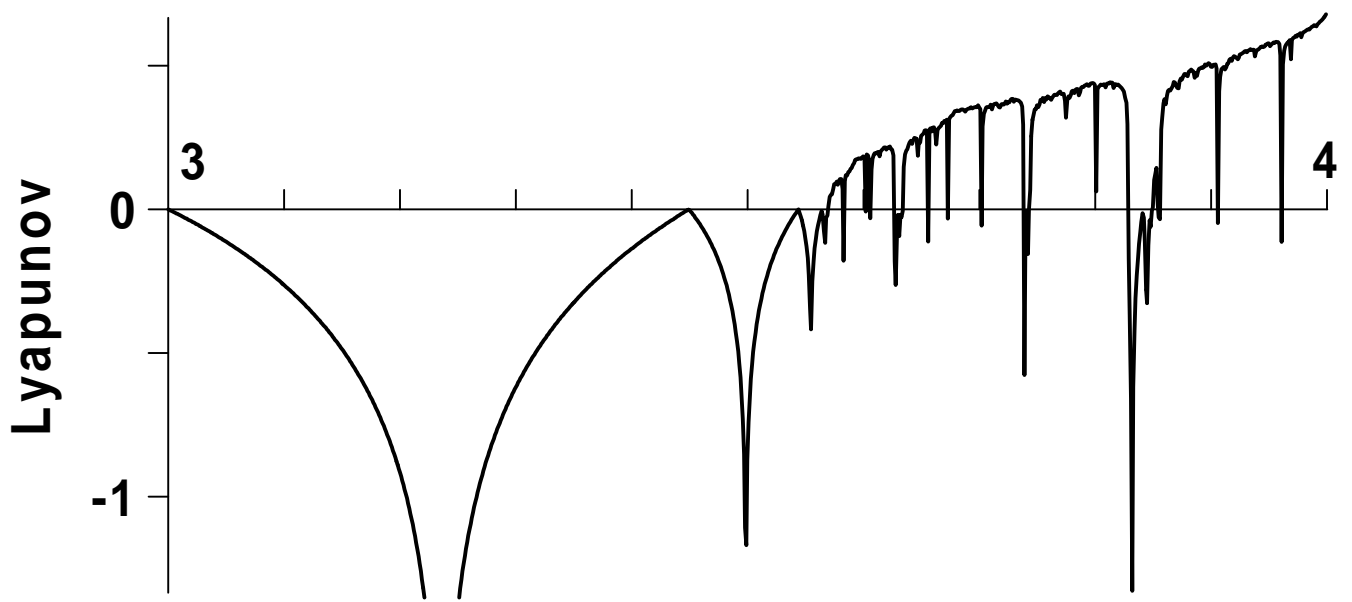
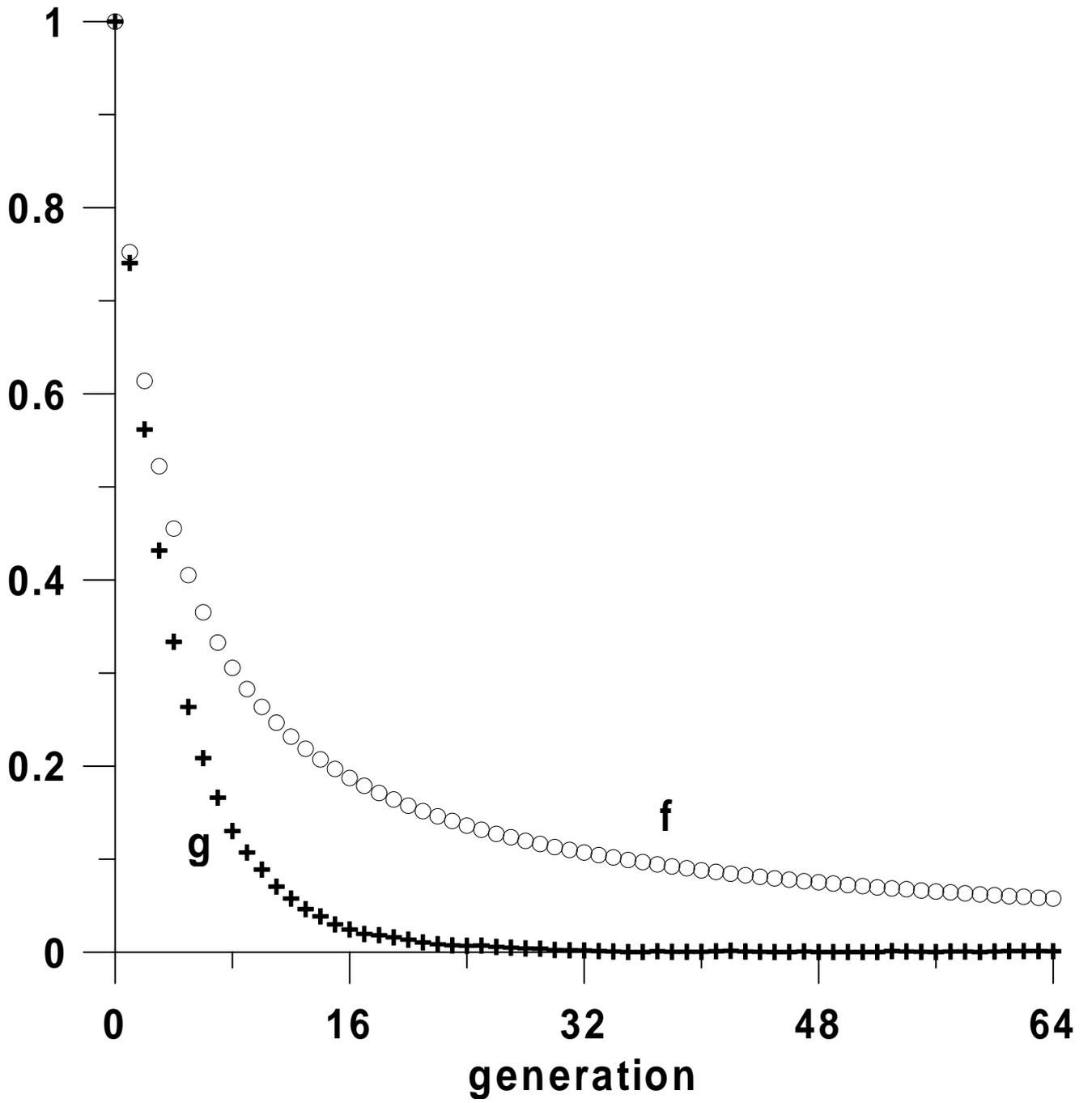
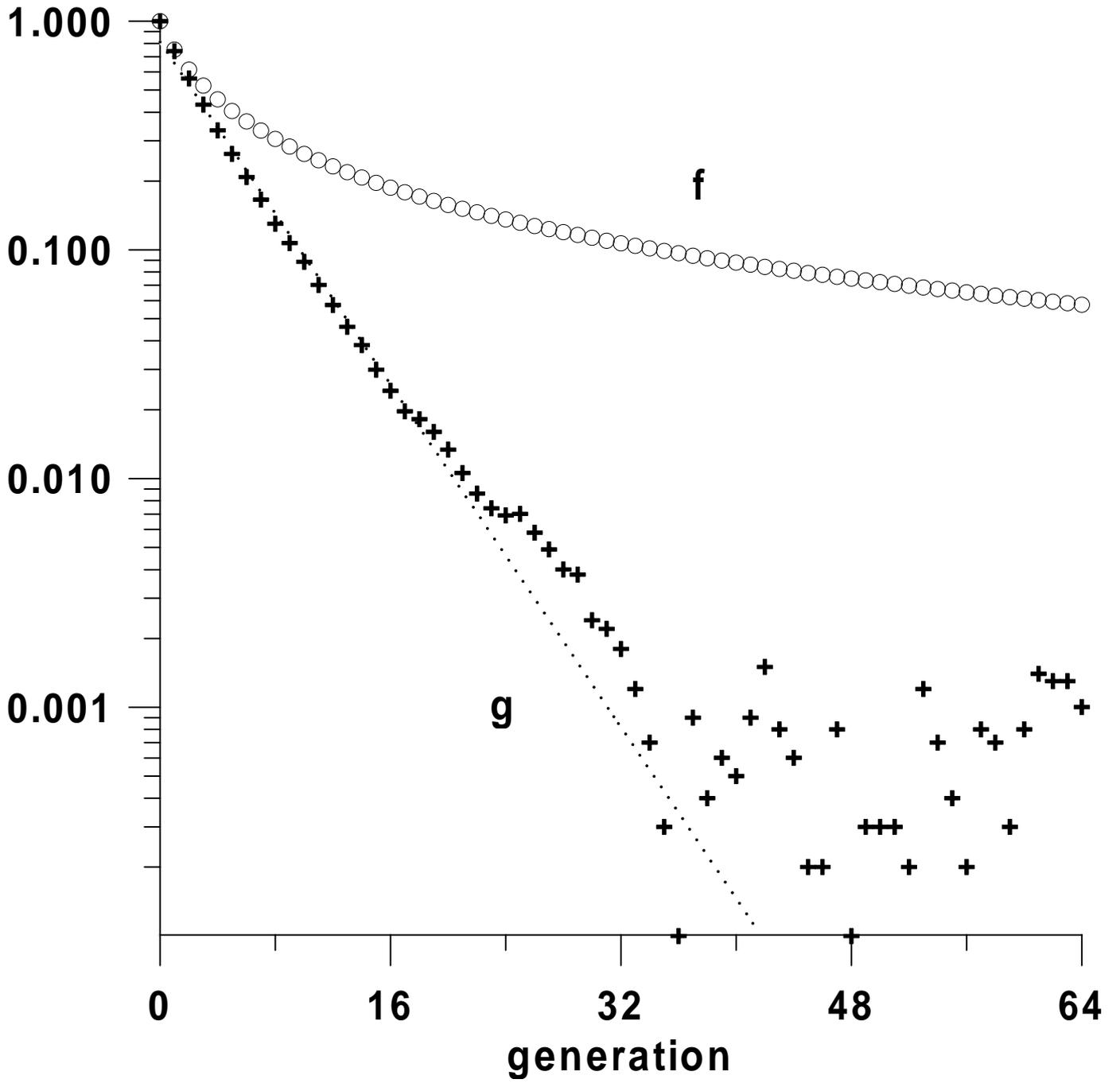
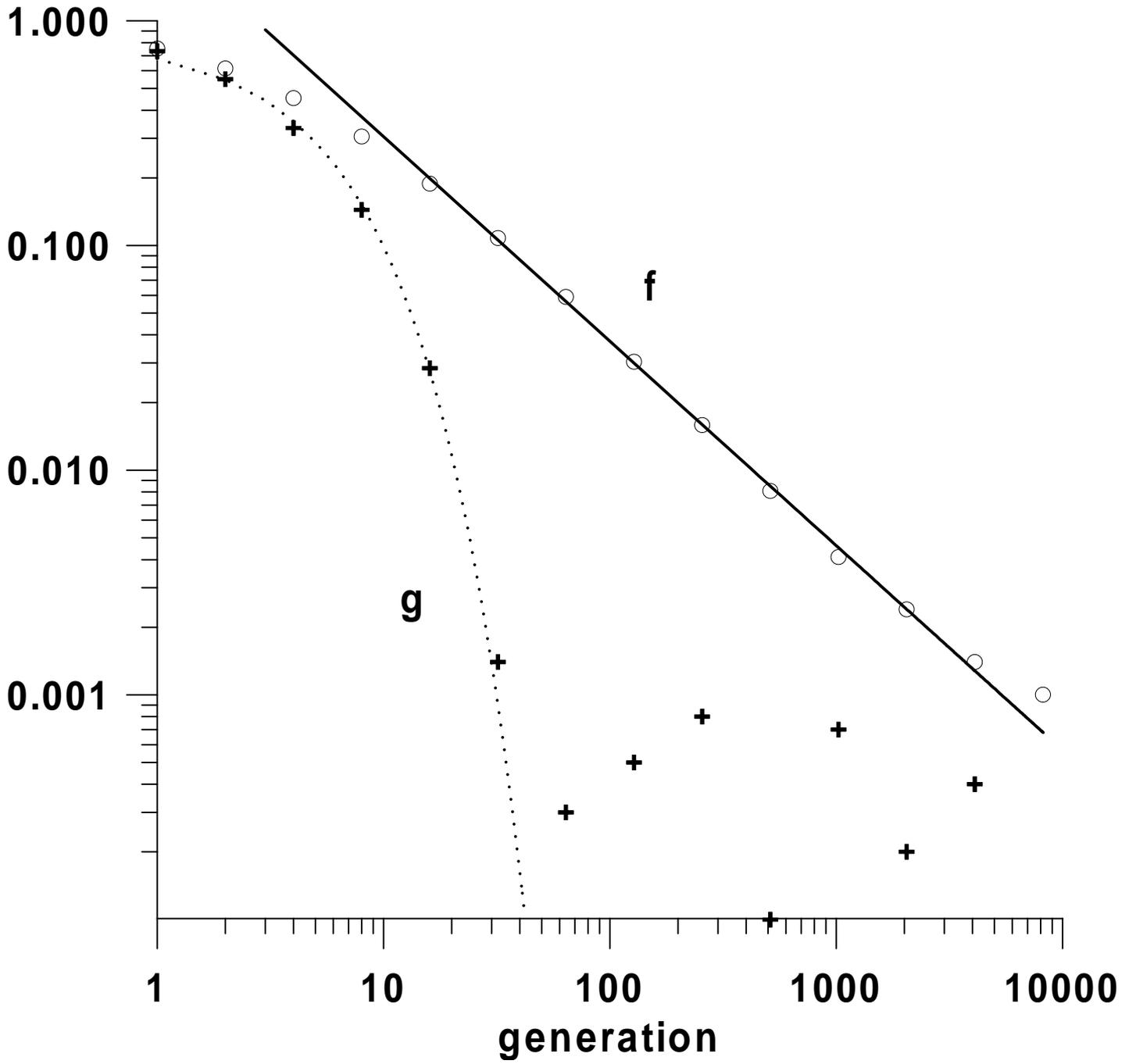
\bye